\documentclass{article}
\usepackage{times}
\usepackage{xcolor}
\usepackage{soul}
\usepackage[utf8]{inputenc}
\usepackage[small]{caption}
\usepackage{enumitem}
\usepackage[colorlinks=true,linkcolor=blue,urlcolor=blue,citecolor=green!65!black]{hyperref}
\usepackage{amsmath}
\usepackage{graphicx}
\usepackage{float}
\usepackage{subcaption}
\usepackage{amsmath}
\usepackage{amsthm, amssymb, amsfonts}
\usepackage{titling}

\pretitle{\begin{flushleft}\LARGE\bfseries}
	\posttitle{\par\end{flushleft}}
\preauthor{\begin{flushleft}\large}
	\postauthor{\par\end{flushleft}}
\predate{\begin{flushleft}}
	\postdate{\par\end{flushleft}\vskip 0.5em}

\newcommand{\RMSE}{\mathrm{RMSE}}
\newcommand{\EE}{\mathbb{E}}

\usepackage{xcite}

\usepackage{xr}
\makeatletter
\newcommand*{\addFileDependency}[1]{% argument=file name and extension
  \typeout{(#1)}% latexmk will find this if $recorder=0 (however, in that case, it will ignore #1 if it is a .aux or .pdf file etc and it exists! if it doesn't exist, it will appear in the list of dependents regardless)
  \@addtofilelist{#1}% if you want it to appear in \listfiles, not really necessary and latexmk doesn't use this
  \IfFileExists{#1}{}{\typeout{No file #1.}}% latexmk will find this message if #1 doesn't exist (yet)
}
\makeatother

\newcommand*{\myexternaldocument}[1]{%
    \externaldocument{#1}%
    \addFileDependency{#1.tex}%
    \addFileDependency{#1.aux}%
}
\myexternaldocument{Supplementary}

\title{Improving Prediction of Phenotypic Drug Response on Cancer Cell Lines Using Deep Convolutional Network}

\author{
Pengfei Liu$^{1*}$, 
Hongjian Li$^{2,3}$,
Shuai Li$^1$,
Kwong-Sak Leung$^1$, 
\\ 
$^1$ Department of Computer Science and Engineering\\
The Chinese University of Hong Kong, Sha Tin, N.T., Hong Kong.\\
$^2$ SDIVF R\&D Centre, Hong Kong Science Park \\
Sha Tin, N.T., Hong Kong.\\
$^3$ School of Biomedical Sciences \\
The Chinese University of Hong Kong, Sha Tin, N.T.    , Hong Kong.\\
\{pfliu, shuaili, ksleung\}@cse.cuhk.edu.hk, jackyleehongjian@gmail.com\\
$^*$Correspondence: pfliu@cse.cuhk.edu.hk
}
\begin{document}
\maketitle

\begin{abstract}

Understanding the phenotypic drug response on cancer cell lines plays a vital rule in anti-cancer drug discovery and re-purposing. The Genomics of Drug Sensitivity in Cancer (GDSC) database provides open data for researchers in phenotypic screening to test their models and methods. Previously, most research in these areas starts from the fingerprints or features of drugs, instead of their structures. In this paper, 
we introduce a model for phenotypic screening, which is called twin Convolutional Neural Network for drugs in SMILES format (tCNNS). tCNNS is comprised of CNN input channels for drugs in SMILES format and cancer cell lines respectively. Our model achieves $0.84$ for the coefficient of determinant($R^2$) and $0.92$ for Pearson correlation($R_p$), which are significantly better than previous works\cite{ammad2014integrative,haider2015copula,menden2013machine}. Besides these statistical metrics, tCNNS also provides some insights into phenotypic screening.
%Describe the revival of phenotypic screening and relevant datasets. Introduce related works and our motivation. Present our methods and results. Make conclusions.
\end{abstract}

\section{Introduction}
\label{sec:introduction}

% more detailed introduction to Phenotypic screening
Historically, drug discovery was phenotypic by nature. Small organic molecules exhibiting observable phenotypic activity (e.g. whole-cell activity) were detected, with a famous example being penicillin, which was serendipitously found. Phenotypic screening, an original drug screening paradigm, is now gaining new attention given the fact that in recent years the number of approved drugs discovered through phenotypic screens has exceeded those discovered through molecular target-based approaches.  The latter, despite being the main drug discovery paradigm in the past 25 years, can potentially suffer from the failure in identifying and validating the therapeutic targets. In reality, most FDA approvals of first-in-class drugs actually originated from phenotypic screening before their precise mechanisms of actions or molecular targets ware elaborated. A popular example of this is aspirin (acetylsalicylic acid), for which it took nearly a century to elucidate the mechanism of its actions and molecular targets.

%%GDSC com from Phenotypic screening
%%Introduce CCLE and GDSC. Describe the motivation of CCLE and GDSC. Why did people conduct drug-cell experiments systematically and collect them into databases like CCLE and GDSC? It is because the revival of phenotypic screening.
There are some public phenotypic screening datasets online to support the study the pharmacologic functions of drugs. Cancer Cell Line Encyclopedia (CCLE) and Genomics of Drug Sensitivity in Cancer (GDSC) are the most popular datasets in the field\cite{cancer2015pharmacogenomic}.

%% related work on GDSC
A pioneer work using machine-learning approaches to predict drug response on cancer cell lines was by Menden et al. \cite{menden2013machine}. The authors used a neural network to analyze the response of drugs to cancer cell lines on the GDSC dataset. Their main result was the achievement of $0.72$ for the coefficient of determination and $0.85$ for Pearson correlation. \cite{ammad2014integrative} and \cite{haider2015copula} are other two works on GDSC dataset. The first used kernelized Bayesian matrix factorization to conduct QSAR analysis on cancer cell lines and anti-cancer drugs, and the second used multivariate random forests. Both of their results are not as good as those in \cite{menden2013machine}, which is chosen to be the benchmark for our work.

%% add something connect machine learning to deep learning
The first wave of applications of deep learning in pharmaceutical research has emerged in recent years. Its utility has gone beyond bioactivity predictions and has shown promise in addressing diverse problems in drug discovery. Examples cover bioactivity prediction\cite{mitchell2014machine}, de novo molecular design\cite{goh2017deep}, synthesis prediction\cite{mamoshina2016applications} and biological image analysis\cite{cruz2013deep,litjens2017survey}. A typical example of applying deep learning in protein-ligand interaction is the investigation done by Ragoza et al\cite{ragoza2017protein}.

Convolutional neural network (CNN) is a model in machine learning which can detect the patterns in data and support classification and regression\cite{yosinski2014transferable}. CNN has achieved breaking-through results in many areas including pharmaceutical research \cite{kalchbrenner2014convolutional,wang2016protein,mobadersany2018predicting} and has won the championship in ImageNet-2012\cite{krizhevsky2012imagenet}.

%%AtomNet: a deep convolutional neural network for bioactivity prediction in structure-based drug discovery \cite{wallach2015atomnet} showed that it is possible apply CNN on molecular 3D structure directly. Atomic Convolutional Networks for Predicting Protein-Ligand Binding Affinity \cite{gomes2017atomic} .

Inspired by the achievements of CNN in these areas, we are interested to see if CNN, compared to conventional machine-learning techniques\cite{menden2013machine,ammad2014integrative,haider2015copula}, could significantly improve the prediction accuracy of phenotypic drug response on cancer cell lines. CNN has not been applied to this problem. This is the first study in exploiting deep convolutional neural networks in predicting phenotypic anti-cancer drug response.

In this paper, we introduce a twin CNN networks model applied to drugs in SMILES format as input, and we call this model tCNNS. tCNNS comprises a CNN network for drugs and another CNN network for cancer cell lines, which will be explained in detail later. Comparing to previous work, we use the latest version of the GDSC dataset, which is bigger and more complete, and our model is more advanced. Most importantly, we achieve much better results than previous works. We share our model online, hoping to make a contribution to other researchers.
\section{Relate Work}
\label{sec:relate_work}

%%\shuai{In summary, the logic is not clear in this section. You should think a flow first on how you want to develop this part.}

Recently, in drug discovery, researchers start to use the molecular structure of drugs directly as features \cite{gomez2016automatic,gomes2017atomic,gomez2018automatic,altae2017low} instead of using extracted features from open source softwares\cite{czarnecki2015weighted,vass2016molecular}. Due to their good ability to process high-dimensional structure on data, deep learning has been largely adopted in this area\cite{lavecchia2015machine,sliwoski2014computational,gawehn2016deep}.

In \cite{gawehn2016deep}, the author stated that both the qualitative classifiers and the quantitative structure-activity relationship (QSAR) models in this area depend on the molecular descriptor, which is \textit{the decisive step in the model development process}. In the recent two years, there are several different deep neural network models that are trained directly from drugs structure and avoiding the \textit{decisive step}. Those deep neural network models include unsupervised auto-encoder (AE), and supervised convolution neural network(CNN), recurrent neural network (RNN).

\cite{gomez2016automatic} and \cite{kearnes2016molecular} converted  simplified molecular input line entry specification (SMILES) of drug into vectors using unsupervised auto-encoder. Those vectors can be used as features or fingerprint of drugs. \cite{altae2017low} and \cite{xu2017seq2seq} further extended this method for drug discovery. They predicted the use of drugs by comparing the similarity between those vectors of drugs. 

One of the direct ways to apply CNN on drug structures is to apply CNN on the image of drugs\cite{goh2017much}, instead of the formulas of drugs. In \cite{goh2017much}, the author adopts a computer vision method to screen the image of drugs. The advantage of starting from the image of drugs instead of the formulas of drugs is that it can avoid the massive work of handling the diversity of drugs. However, the disadvantages are that the accuracy is compromised because the information will be distorted when mapping drugs structures to images and the performance of this method relies on the quality of image processing.

Beside applying CNN on drug images, it is also possible to apply CNN on molecular 3D structures directly\cite{wallach2015atomnet}. In this paper,  the author predicted the binding energy of the small area around an atom, instead of the entire structure of drugs. We think that it will be interesting to compare the different representations of drugs, such as the 3D structured, the feature vectors learned from SMILES and the features extracted from other software like PaDEL \cite{yap2011padel}. They may have different influences on different problems. However, we did not find any kind of such comparisons.

Even though RNN is usually used to handle time sequence data\cite{yao2015describing} instead of spatial data, it is very impressive that \cite{lusci2013deep} applied RNN on molecule SMILES to predict its solubility. They first converted the SMILES into indirect graphs, and then fed it into an RNN. In this work, they only considered the property of drugs alone, without considering the interactions among drugs and other biological factors, such as cell lines or proteins.

The benchmark work we used to compare with our model is \cite{menden2013machine}, where the author used a neural network to analyze the $\text{IC}_{50}$ of drugs to cancer cells on the same dataset as ours. We think their network structure is not advanced enough, and the features they used are not informative enough. We design our tCNNS, a convolution neural network (CNN) based model, to predict the interaction between drugs and cell lines, and it performs better in the experiments. 
\section{Methods}
\label{sec:model}

In this section, we will introduce the chosen database GDSC with preprocessing steps and the adopted neural network structure in details to make our experiments easier to replicate.

%Use passive voice in Methods.

\subsection{Data Acquisition and Preprocessing}

Genomics of Drugs Sensitivity in Cancer (GDSC) \cite{garnett2012systematic} is a public online database about the relationship among many types of cancer and various anti-cancer drugs. Cancer cell lines in GDSC are described by their genetic features, such as mutations state and copy number variances. For the drugs, GDSC provides their names and the compound id (CID). In chemistry, CID is a unique number assigned to each molecular and can be used as the reference number to extract more information about the drugs such as their molecular structures from other databases. GDSC uses $\text{IC}_{50}$ as the metric of drugs' effect on cancers, which means how much of the drugs is needed to inhibit cancer by half. The less the value, the more effective the drug is. GDSC is an ongoing project and is being updated regularly. In this paper, we use version $6.0$ of GDSC, and the benchmark of our paper \cite{menden2013machine} used version $2.0$ of the GDSC.

We downloaded three files from GDSC:
\begin{enumerate}[label=(\alph*)]
	\item Drug$\_$list.csv, which is a list of $265$ drugs. Each drug can be referred by its CID or name.
	\item PANCANCER$\_$Genetic$\_$feature.csv, which is a list of $990$ cancer cell lines from $23$ different types of cancers. Each cell line is described by at most $735$ features. Any feature belongs to one of the two categories: mutation state or copy number alteration.
	\item PANCANCER$\_$IC.csv, which contains the $\text{IC}_{50}$ information between $250$ drugs and $1074$ cell lines.
\end{enumerate}

%% 310 mut, 425 cna here
%% motivation, short coming of the dataset
Note that the numbers of drugs in the file (a) and the file (c) are inconsistent, and the numbers of cell lines in the file (b) and the file (c) are also inconsistent. Some cell lines have less than $735$ features. Besides, GDSC does not provide the features for drugs, which have to be downloaded from other datasets. All of these indicate that a preprocessing is needed to clean the data. 

%%Always describe the motivation of any data cleansing operation before we describe the operation itself.

The first step is to purify the drug list. There are $15$ repeating items in the file (a) which are removed. Some CIDs in the file (a) are inconsistent with the CIDs that in PubChem \cite{kim2015pubchem}, which is a popular public chemical compounds database. To keep the consistency, we use the CIDs from PubChem. Some drugs cannot be found in PubChem by referring their names in the file (a) and they are removed. In the end,  $223$ drugs with both names and CIDs are left.

The second step is to purify the cell lines list. For the $990$ cell lines in the file (b), $42$ of them has less than $735$ features. We remove these cell lines and there are $948$ cell lines left.

%%Explain what mutation states are, how they are detected, and how they are represented (in the form of 0 or 1). Perhaps give an example of mutation state. 

%%Explain what IC50 is, and how the values are represented (in natural logarithmic scale). Describe the min and max concentration tested. Explain any value between [min, max] is reliable and those outside the range was obtained from extrapolation, which are not as reliable. Give figures to show number of IC50 inside and outside range.
%% draw a multi-dimenion figure, or statisitics

%% give the percentage of the data
In the third step, we only keep the $\text{IC}_{50}$ values between the remaining drugs after the first step and the remaining cell lines after the second steps.  All the other $\text{IC}_{50}$ values in the file (c) are removed.
In summary, there are $223$ drugs and $948$ cell lines after the preprocessing. Amon the g $223 \times 948 = 211,404$ interacting pairs, $81.4\%$ ($172,114$) of the $\text{IC}_{50}$ values are provided in file (c), and $18.6\%$ ($39,290$) of the $\text{IC}_{50}$ are missing.

The $\text{IC}_{50}$ data in file (c) are the logarithm of their real value. To make it easy for training and comparison, we follow the method in \cite{menden2013machine} to normalize the logarithmic $\text{IC}_{50}$ values into the $(0, 1)$ interval. Given a logarithmic $\text{IC}_{50}$ value $x$, we first take exponential of it to get the real value $y = e^x$ and then use the following function to normalize $y$: 
$$
y \mapsto \frac{1}{1 + y^{-0.1}}\,.
$$
Usually $y$ is very small ($<10^{-3}$),  and the parameter $-0.1$ is chosen to distribute the result more uniformly on the interval $(0,1)$ \cite{menden2013machine}.

\subsection{Numerical Descriptor Extraction}

%Describe drug features. PaDAL features. what are they. how many. compare to menden2013machine and related works.

%%Describe cell features. The cells are converted into binary vectors with length $735$.
%% detailed explain of vectorize

%% add citation, not novel
Recently, there are some pioneering works that apply deep neural network (DNN) directly on the simplified molecular-input line-entry system (SMILES) of drugs. SMILES is a linear notation form to represent the structure of molecules, in which letters, digits and special characters are used to represent the chemical elements in a molecule. For example, "\texttt{C}" is for carbon atom, "\texttt{=}" is for covalent bond between two atoms, Carbon dioxide can be represent as \texttt{O=C=O}, and aspirin can be represented as \texttt{O=C(C)OC1CCCCC1C(=O)O}.

There are some challenges to apply convolution neural network (CNN) on drugs in SMILES format: first, SMILES can be constructed in various ways and there can be many possible SMILESs for each drug;  second, the size of the samples for a CNN should be consistent, but the lengths the SMILES format of drugs are different from each other; third, and more importantly, the SMILES format is composed of different letters representing different chemical elements, such as atoms and bonds, and it does not make sense to apply convolution operation among different chemical elements. To solve these problems, preprocessing is needed to convert the SMILES into a uniform format, and restrict convolutional operation among the same chemical elements.

To keep unique SMILES format for the drugs, we use the \textit{canonical SMILES}\cite{o2012towards} for the drugs. Among $223$ drugs, the canonical SMILES of $184$ of them can be found from PubChemPy by the drugs names, which is a python interface for PubChem. The canonical SMILES of the remaining $39$ drugs are downloaded from the Library of the Integrated Network-based Cellular Signatures (LINCS) \cite{keenan2017library} instead.

%% draw graph of distribution

The longest SMILES for the drugs contains $188$ symbols, and most SMILES lengths are between $20$ and $90$. To keep the size consistent and retain the complete information, all the SMILESs are left aligned with space padding in the right if they are shorter than $188$. 

The neural network cannot directly take the drugs in SMILES format as input, and we need to convert the SMILES format (they are uniform length now after handling the second challenge) into a format that can be used in the neural network. There are $72$ different symbols in the SMILES format for the total $223$ drugs. The distribution of these symbols is quite imbalanced. For example, carbon atom \texttt{[C]} appears in all the $223$ drugs, but both the gold atom \texttt{[Au]} and chlorine atom \texttt{[Cl]} appear only once. Suppose we use rows to represent different symbols, and use columns to represent positions in the SMILES format, then each drug in SMILES format can be converted into a $72*188$ one-hot matrix which only contains $0$ and $1$. The row number $72$ represents $72$ possible symbols, and column number $188$ represents the longest length of SMILES. In the one-hot matrix for a drug, a value $1$ at row $i$ and column $j$ means that the $i$th symbol appears at $j$th position in the SMILES format for the drug. In our model, the column of the one-hot matrix is treated as different channels in CNN, and the $1$D convolutional operation will be applied along each row of the one-hot matrix, which restricts convolutional operation within the same chemical element.

\subsection{Deep Neural Network}

\begin{figure*}[h]
	\caption{The upper part is the branch for drugs, and the lower part is the branch for cell lines. Both of them are connected to a fully connected network in the right. The general work-flow of our model is from left to right. The left is the input data of one-hot representations for drugs and the feature vectors for cell lines. The black square stands for $1$ and empty square stands for $0$. In the middle is the two convolution neural networks for drugs and cell lines. They take the one-hot representations and feature vectors as input data respectively, and their outputs can be interpreted as the abstract features for drugs and cell lines. The structures of the two convolution neural networks are similar. The right is a fully connected network that does regression analysis from the $\text{IC}_{50}$ to the abstract features from the two CNNs in the middle part.} 
	\label{fig:Architecture}
\end{figure*}
%\shuai{The middle part is bad in this figure. Borrow how people draw cnn-pooling-relu from vision paper.} %\shuai{I guess you should rotate the input so that the horizontal direction shows the channel}

The structure of our model is shown in Figure \ref{fig:Architecture}. The input data for our model is the one-hot representation of drugs (phenanthroline is used as an example for the drugs) and the feature vectors of the cell lines. The work-flow can be divided into two stages as follows.

%% explain the everything in the figure
During the first stage, we build a model with two convolutional neural networks (CNN) branches to distill features for drugs and cell lines separately. We use $1$D CNN for the cell-line branch since the input data is $1$D feature vectors for cell lines. We also use $1$D CNN for the drug branch and treat different symbols as different channels in CNN. The convolution is applied along the length of the SMILES format. The structures for the two branches are the same. For each branch, we use three similar layers: each layer with convolution width $7$, convolution stride $1$, max pooling width $3$, and pooling stride $3$. The only difference among the layers is that their number of channels is $40$, $80$ and $60$, respectively. The choices of those parameters for the CNN are inspired by the model in \cite{kelley2016basset}, in which the author chose a three-layers network model and used a prime number as the filter width. We find that either reducing the pooling size or adding the channel number has the potential to enhance our model but with the cost of losing stability. Losing stability means that the experiments results sometimes become un-repeatable and it will be explained in detail in Section \ref{sec:results}.

In the second stage, after the two branches of CNN, there is a fully connected network (FCN), which aims to do the regression analysis between the output of the two branches and the $\text{IC}_{50}$ values. There are three hidden layers in the FCN, each with $1024$ neurons. We set the dropout probability to be $0.5$ for the FCN during the training phase \cite{kelley2016basset}.

%% tensorflow
We implemented our model on TensorFlow v$1.4.0$ \cite{abadi2016tensorflow}, which is a popular DNN library with many successful applications \cite{sawant2018brain,abadi2016tensorflow}. The data and code of this paper can be downloaded from \url{https://github.com/Lowpassfilter/tCNNS-Project}.

\subsection{Performance Measures}

%%Define R2, Rp, RMSR. Explain their semantics.
We adopt the metrics of the coefficient of determination ($R^2$), Pearson correlation coefficient ($R_p$), and root mean square error ($\RMSE$) to measure the performance of our model, as in the benchmark paper \cite{menden2013machine}.

$R^2$ measures variance proportion of the dependent variables that is predictable from the independent variables. Let $y_i$ be the label of a sample $x_i$, and our label prediction on $x_i$ is $f_i$. The error of our prediction, or residual, is defined as $e_i = y_i - f_i$. Let the mean of $y_i$ be $\bar{y} = \frac{1}{n} \sum _i y_i$, We have total sum of squares:

$$SS_{\mathrm{tot}} = \sum_i (y_i - \bar{y})^2$$
regression sum of squares:
$$SS_{\mathrm{reg}} = \sum_i (f_i - \bar{y})^2$$
residual sum of squares:
$$SS_{\mathrm{res}}=\sum_i (y_i - f_i)^2 = \sum_i e_i^2$$
$R^2$ is defined as:
$$R^2 = 1 - \frac{SS_{\mathrm{res}}}{SS_{\mathrm{tot}}}.$$

$R_p$ measures the linear correlation between two variables. We use $Y$ as the true label and $F$ as the corresponding prediction for any sample. Let the mean and standard deviation of $Y$ be $\bar{Y}$ and $\sigma_Y$ respectively, and those for the prediction $F$ be $\bar{F}$ and $\sigma_F$ respectively. $R_p$ is defined as:
$$
R_p = \frac{\EE[(Y - \bar{Y})(F - \bar{F})]}{\sigma_Y\sigma_F} \,.
$$

$\RMSE$ measures the difference between two variables $Y$ and $F$, and $\RMSE$ is defined as:
$$
\RMSE = \sqrt{\EE[(Y-F)^2]} \,.
$$

\section{Results and Discussion}
\label{sec:results}
In this section, we demonstrate the performance of our algorithm tCNNS with various data input settings, and it achieves high scores in the circumstances we set in the experiments. 

\subsection{Rediscovering Known Drug-Cell Line Responses}

In the $223 \times 948$ drug-cell line interaction pairs, GDSC provides the $\text{IC}_{50}$ for $172,114$ of them. In this part, we split those known pairs into $80\%$ as the training set, $10\%$ as the validation set, and $10\%$ as the testing set. In each epoch, parameters in tCNNS are updated using gradient descent on the training set. The validation set is used to check the performance of the tCNNS. If the $\RMSE$ on the validation set does not decrease in $10$ recent epochs, the training process stops and the predictions of our model on the testing set are compared with the given $\text{IC}_{50}$ values in GDSC. 

The experiments are set in this way to stimulate the real situations in which we want to train a model on data with labels and apply the model on data without labels. The validation set is separated from the training set so that we can choose a suitable time to stop training independent and avoid the problem of over-fitting.

We check the regression accuracy of our model by comparing our predictions against the $\text{IC}_{50}$ provided in GDSC, and the results are displayed in Figure \ref{fig:regression}.

\begin{figure}[h]
	\centering
	\caption{Regression results on testing set compared to the ground truth $\text{IC}_{50}$ values. The $x$ axis is the experimental $\text{IC}_{50}$ in natural logarithmic scale, and the $y$ axis is the predicted $\text{IC}_{50}$ in natural logarithmic scale. Different colors demonstrate how many testing samples fall in each small square of $0.1 \times 0.1$, or the hot map of the distribution, where dark purple indicates more samples (around $30$ samples per small square $0.1 \times 0.1$) and light blue indicates fewer samples (less than $5$ samples per small square $0.1 \times 0.1$)}
	\label{fig:regression}
\end{figure}

%# Here we present three figures: 1) regression result using drug features from PaDAL only. 2) regression result using drug features from SMILES only. 3) regression result using drug features from both PaDAL and SMILES.
%# In addition to R2 and Rp, also create new plots to show RMSR.

In the experiments, our model tCNNS outperforms previous work \cite{menden2013machine} in many ways: The efficiency of determination ($R^2$) is increased from $0.72$ to $0.84$, the Pearson correlation($R_p$) is increased from $0.85$ to $0.92$, and the root mean squared error ($\RMSE$) is reduced from $0.83$ to $0.027$. 

%% discussion
Many hyper-parameters affect the performance of our model tCNNS, such as the number of layers and filter size. After examining, we find that smaller pooling size and more numbers of channels can enhance the performance furthermore but decrease the stability. Take the pooling size for example. When the it is reduced from $3$ to $2$, $R^2$ is further increased from $0.84$ to $0.92$ and $R_p$ is further increased from $0.93$ to $0.96$. The cost of this enhancement is that the network becomes unstable and diverge \cite{kawaguchi2016deep} during the training. To keep the experiments results repeatable, we report the results with parameters that ensure our experiments stable

In Figure \ref{fig:regression}, we notice that tCNNS is most accurate in the middle part, but less accurate in the two ends in the figure. Its outputs are not small enough when the experimental $\text{IC}_{50}$ values are very small and are not big enough when the data are huge. This means that tCNNS can be further optimized if we can enhance its performance in the two ends.

%%Compare to previous works. menden2013machine used v2.0, whose cell features contained not only XXX mutation states but also copy number variation and micro satellite.
In the experiment, we do not recover the performance reported in \cite{menden2013machine} using the network structure in it.  After replacing its network with tCNNS, We find that tCNNS do not converge using the features extracted from PaDEL. Finally, 
we replace the network in \cite{menden2013machine} with a deeper one, a network with three hidden layers and $1024$ neurons in each hidden layer. This modified benchmark gets the $R^2$ around $0.65$ and $R_p$ around $0.81$, which is shown on supplementary Figure 1. From this figure, we can see that the result is clearly horizontally stratified, which means that the neural network lacks representation power using PaDEL features.

Based on these results, we conclude that the connections among the $\text{IC}_{50}$ values and the SMILES format of drugs are stronger than that among the $\text{IC}_{50}$ values and features extracted using PaDEL. 

\subsection{Predicting Unknown Drug-Cell Line Responses}

Previously, we split the known drug-cell line interaction pairs into training, validation, and testing set to check the accuracy of tCNNS. In this part, we predict the values for those missing pairs by training tCNNS on the known items. The known pairs are split into $90\%$ as the training set, and $10\%$ as the validation set. Again, if the $\RMSE$ on the validation set does not decrease in $10$ recent epochs, the training process stops and the trained network will be used to predict the values for the missing items. The results are shown in Figure \ref{fig:prediction}.

\begin{figure*}[!h]
	\centering
	\caption{The predicted missing $\text{IC}_{50}$ values. We range the drugs according to the median of their predicted $\text{IC}_{50}$ values with cells. The horizontal axis is the drug names, and the vertical axis is their negative $log_{10}(\text{IC}_{50})$ values with cells. The left part is the top $20$ drugs with lowest $\text{IC}_{50}$ median, which means that they are probably the most effective drugs, and the right part is the last $20$ drugs with the highest $\text{IC}_{50}$ median, which means that they are the most ineffective drugs. The numbers in each column are the number of missing $\text{IC}_{50}$ values for each drug in the dataset, which can indicate the confidence level of the distribution.}
	\label{fig:prediction}
\end{figure*}
%% more confident with more data
%% explain for outlier

%% professional and detailed discussion
Figure \ref{fig:prediction} is the box plot of the predicted $\text{IC}_{50}$ values for missing items grouped by drugs. For each drug, the box represents the distribution of the values with its related cell lines. The drugs are sorted by the median of the distribution: the $20$ drugs with highest median and $20$ drugs with the lowest median value are plotted. Since we do not know the real value for these missing pairs, the accuracy of our prediction is stated by survey and analysis as follows. 

\textit{Bortezomib} is the best drug in our prediction. In fact, the top $40$ pairs with the lowest $\text{IC}_{50}$ value are all from \textit{Bortezomib} with some other cell lines. The outstanding performance of \textit{Bortesomib} in missing pairs is consistent with that in the existing pairs. There is some supporting information in \cite{friedman2015landscape} that the author found that \textit{Bortezomib} can sensitize cell lines to many other anti-cancer drugs. 

%\pfliu{I should add an exemple here to show how this drug can sensitize other drugs.}

\textit{Aica ribonucleotide} and \textit{Phenformin} have the poorest performance in our prediction. Based on our survey, the former one is initially invented to prevent blood flow, and the later one is initially used as an anti-diabetic drug. These two drugs have the potential to cure cancer because they can inhibit the growth of cell (\textit{Aica ribonucleotide}) or inhibit the growth of Complex I (\textit{Phenformin}), but their effects are limited since anti-cancer is only the side effect of these them, not their main function.

Based on our predictions, the $\text{IC}_{50}$ of drug \textit{Bortezomib} with cell line \textit{NCI-H2342} is $1.19*10^{-4} \mu g$. This is the third smallest in our predictions and it indicates a good therapeutic effect. This prediction is supported by the findings in \cite{ge2010quantitative}\cite{hornbeck2014phosphositeplus} that \textit{Bortezomib} is able to control Phosphorylation that causes lung cancer and \textit{NCI-H2342} is a lung cell line. Similar evidence to supporting this prediction can also be found in Cell Signaling Technology's 2011 published curation set. \footnote{https://www.phosphosite.org/siteAction.action?id=3131}

\subsection{Retraining Without Extrapolated Activity Data}
%% repeat 3.1 figures.
%%Describe again the min and max concentration tested in GDSC. any value between [min, max] is reliable and those outside the range was obtained from extrapolation, which are not as reliable. 
%%Give figures to show number of IC50 inside and outside range. Hence, we repeated the experiments with only the data that fall within the range.

%\pfliu{conc data is much faster}
For each drug in GDSC, there are two important thresholds called minimum screening concentration (min{\_}conc), which is the minimum $\text{IC}_{50}$ value verified by biological experiments, and maximum screening concentration (max{\_}conc), which is the maximum $\text{IC}_{50}$ value verified by biological experiments. Any $\text{IC}_{50}$ beyond these two thresholds is extrapolated, and not verified by experiments. In general, $\text{IC}_{50}$ value within min{\_}conc and max{\_}conc is more accurate than those outside of the thresholds.

In the GDSC data that we use, only max{\_}conc is provided, and there are $64,440$ $\text{IC}_{50}$ values below  max{\_}conc, which is about $37\%$ in the whole existing $172,114$ $\text{IC}_{50}$ values. 

In this part, tCNNS is trained on the $\text{IC}_{50}$ values below the max{\_}conc threshold. We keep $10\%$ data for validating and $10\%$ data for testing and reduce the percentage of data used for training from $80\%$ to $1\%$. The regression result is shown in supplementary Figure 2 and the comparison against the tCNNS which trained on whole existing data is showed in Figure \ref{fig:percentage}.
%% computational burden
\begin{figure}[h]
	\centering
	\caption{The performance with different percentages of data used. The x-axis is the percentage of data we use as training data from the total existing $\text{IC}_{50}$ values ($172114$) in the database. Since we will use $10\%$ for validating and $10\%$ for testing, the max x is $80\%$. The y-axis is the performance of our model. The solid lines represent the result on total existing data, and the dash lines represent the results where we just use the $\text{IC}_{50}$ values below the max screening concentration threshold(max{\_}conc), below which the data is more accurate. Since there are only $64,440$ values below max{\_}conc, so the dash lines end at around $\frac{64,440}{172,114}*80\% = 30\%$}
	\label{fig:percentage}
\end{figure}
From supplementary Figure 2, we see that tCNNS can achieve almost the same good result on max{\_}conc data only, which is faster because less data is needed. There are some other properties of tCNNS that we can conclude from Figure \ref{fig:percentage}. Firstly, it performs very well even with very limited training data. For example, even trained on $1\%$ of the existing $\text{IC}_{50}$ values, $R^2$ can be almost $0.5$ and $R_p$ be around $0.7$. Secondly, and more importantly, tCNNS performed better with less and more accurate data. We see that the dash lines (results on data below max$\_$conc) are always above the solid lines (result on all data), and the final performance on max$\_$conc data is almost as good as that on the total data, although the amount of data for the former is only $37\%$ of the later. To further compare the best performance on all data and max$\_$conc data only, we repeated the experiments for 20 times to see the distribution of the results, and show it in supplementary Figure 3.

%% definition of random, relationship to previous figure, zoom in

There are three experimental results shown in Supplementary Figure \ref{fig:max_conc},  which are the experiment on all data, on the data below max$\_$conc and on a random subset of all data with the same size as those below max$\_$conc. Comparing the result on data below max$\_$conc with the result on the random data with the same size, we see that the performance of tCNNS is significantly better on data below max$\_$conc than on random data with the same size, and this proved that tCNNS is able to utilize the accurate data.

Comparing the result on data below max$\_$conc with that on all data, we see that the means of the tow $R^2$ are almost the same, and the $R_p$ on all data is only a little bit better than that on the data below mac$\_$conc. Meanwhile, the variations of $R^2$ and $R_p$ on data below max$\_$conc is a little bigger than those on all data. To conclude, the contribution from low quality extrapolated data is limited, and they can only reduce variation and improve $R_p$ a little bit.

Although the extrapolated data cannot enhance the accuracy of the model, they can help to improve the scalability. The tCNNS model trained on all data performs well when tested on data below max{\_}conc, which is natural because the later is a subset of the former. However, the model trained on data below max{\_}conc performs poorly when tested on all data. $R^2$ drops to $0.33$ and $R_p$ drops to $0.6$. Our explanation is that when tCNNS is trained on all data, it learns some general pattern among all data. On the contrary, when tCNNS is trained on data below max{\_}conc, it only learns the specific patterns for this subset. So although the performance of tCNNS on all data and data below max{\_}conc is similar, the paths they achieve the performance are different.

\subsection{Blind Test For Drugs And Cell lines}

%%Row-wise and column-wise training and test. Perhaps draw a figure to illustrate the concept of drug-blind and cell-blind.

In previous experiments, interaction pairs among drugs and cell lines are randomly selected to be in training set, validation set, or testing set, which means that a specific drug or a specific cell line can exist in training and testing at the same time, which is a problem of drugs re-positioning. In the real world, drugs discovery is a more challenging task in which we may not be able to train tCNNS on a specific drug or cell line before testing on them. Instead, we may need to predict the performance of the specific drug (or cell line) using tCNNS trained on other drugs(or other cell lines.) To stimulate this condition, we design the blind test for drugs and cell lines respectively. 

In the blind test for drugs, drugs are constrained from existing in training and testing at the same time. We randomly select $10\%$ ($23/223$) drugs  and keep their related $text{IC}_{50}$ values for testing. For the remaining $90\%$ drugs, we divide $90\%$ of their related $\text{IC}_{50}$ values for training and $10\%$ for validating. 

In the blind test for cell lines, cell lines are prevented from existing in training set and testing set at the same time. Similar to the case of drugs, we randomly select $10\%$ ($94/948$) cell lines and keep their related $text{IC}_{50}$ values for testing. For the remaining $90\%$ ($904/948$) cells, we used $90\%$ of the related $\text{IC}_{50}$ for training and $10\%$ for testing.

We repeat the blind test for drugs on all data and on the data below max$\_$conc for $150$ times respectively to see the distribution of the results. The same number of experiments for the cell lines are also conducted. The results on all data are shown in Figure \ref{fig:blind}. The results on data below max{\_}conc data are showed on supplementary Figure 4 respectively.

\begin{figure}[h]
	\caption{Drug and cell blind test result on total data. Yellow color boxes represent the result of cells blind, and blue color boxes for drugs blind. From up to down is the result for $R^2$, $R_p$ and RMSR respectively. The red star is the result without controlling data distribution. }
	\label{fig:blind}
\end{figure}

From Figure \ref{fig:blind} and supplementary Figure 4 we see that the performance of tCNNS is comparably robust with the blind test for cell lines, but sensitive with the blind test for drugs. Without the knowledge of drugs in training, the performance drops significantly. Comparing the results in Figure \ref{fig:blind} and supplementary Figure 4, we see that the extrapolated data makes no contribution in this setting.

To conclude, it is very important to have the specific drug or cell line in the training stage before we predict their performance. Experiments results support that even with only one or two related $\text{IC}_{50}$ value, the performance will be significantly improved. For example, \textit{NCI-H378} is a special cell line for lung cancer in GDSC that there are only two $\text{IC}_{50}$ values records for it. For other cell lines, all of them have at least $20$ $\text{IC}_{50}$ values. tCNNS can still make accurate predictions one of the $\text{IC}_{50}$ values for textit{NCI-H378} if the other value is used during the training. Based our the results, the best drug for \textit{NCI-H378}  are \textit{Bortezomib}, which has been explained previously \cite{ge2010quantitative,hornbeck2014phosphositeplus}. 

%\pfliu{I should check these to papers to explain the meaning of this prediction with some details}. 
Meanwhile, tcNNS predicts another potential drug \textit{Docetaxel} for them. The predicted $\text{IC}_{50}$ value between \textit{Docetaxel} and \textit{NCI-H378} is $0.03\mu g$ (third smallest for \textit{NCI-H378}), and the predicted $\text{IC}_{50}$ between \textit{Docetaxel} and \textit{NCI-H250} is $0.04\mu g$ (forth smallest for \textit{NCI-H250}). It is reported in \cite{mohell2015apr} that \textit{APR-246} is a potential useful drug on lung cancer because of its synergy with $TP53$ mutations in lung cells, and there is a \textit{"additive effects"} between \textit{APR-246} and \textit{Docetaxel}.

Comparing the results of the blind test for drugs and blind test for cell lines, we guess the reason why the blind test for cell lines is slightly better is that there is more common information shared among different cell lines and less among drugs. For example, cell lines share similar genetic information, but drugs can be very diversified. To reduce the information sharing among cells lines, we designed another experiment in which cell lines from the same tissue cannot exist in training and testing at the same time. The result is shown in Table \ref{table:tissue_specific}.

In GDSC, the $948$ cell lines belong to $13$ tissue types and $49$ sub-tissue types. We choose to use the $13$ tissue types instead of $49$ sub-tissue types because it can increase the distances and reduce the similarities among different tissues. Each time one tissue type is selected as testing data. For the rest tissues, we mix them together and use $90\%$ for training and $10\%$ data for validation. From Table \ref{table:tissue_specific}, we can see that the performance decrease differently for different tissues. For example, blood has the lowest $R^2$ and $R_p$ in all tissues, which indicates that blood is the most different tissue from other tissues. From Table \ref{table:tissue_specific}, we can also see that the result for validation is much better than testing, which not only because we use validating as the stop criterion for our model, but also because it shares more information with other tissues in the training stage.

The above results in the blind test give us some hint that with limited budgets, we should carefully arrange the \textit{in vivo} experiment to cover a wider range of drugs and cells from different tissues to get better \textit{in silicon} predicting power.

\begin{table}[h]
	\centering
	\begin{tabular}{|c|l|c|c|c|}
		\hline
		Tissue name 			& \begin{tabular}{@{}c@{}}Data \\ Amount\end{tabular} 	&  $R^2$	& $R_p$		& RMSR \\
		\hline
		aero digestive 	& 13806 &0.703 		&0.843		&0.0375      \\
		tract		& 		&(0.826) 	&(0.916)	&(0.0280)      \\
		\hline
		blood			& 31119 &0.500 		&0.724	  	&0.0449      \\
		& 		&(0.833) 	&(0.917)  	&(0.0276)      \\
		\hline
		bone			& 6826  &0.659		&0.813	   	&0.0405     \\
		&		&(0.825) 	&(0.915)   	&(0.0283)      \\
		\hline
		breast			& 9277  &0.657 		&0.811   	&0.0383      \\
		&       &(0.829) 	&(0.919)   	&(0.0281)      \\
		\hline
		digestive 		& 17200 &0.667 		&0.817	   	&0.0384      \\
		system			&       &(0.830) 	&(0.918)   	&(0.0282)      \\
		\hline
		kidney			& 5199  &0.669		&0.819   	&0.0386      \\
		&       &(0.822)	&(0.914)   	&(0.0286)      \\
		\hline
		lung			& 34086 &0.614 		&0.784   	&0.0371      \\
		&       &(0.827) 	&(0.919)   	&(0.0285)      \\
		\hline
		nervous 		& 15763 &0.702 		&0.839   	&0.0364      \\
		system		&       &(0.830) 	&(0.918)   	&(0.0280)      \\
		\hline
		pancreas		& 5358  &0.703 		&0.840   	&0.0370      \\
		&       &(0.820) 	&(0.913)   	&(0.0287)      \\
		\hline
		skin			& 10488 &0.676 		&0.824   	&0.0394      \\
		&       &(0.827) 	&(0.917)   	&(0.0281)      \\
		\hline
		soft tissue		& 3165  &0.712 		&0.853   	&0.0384      \\
		&       &(0.821) 	&(0.914)   	&(0.0284)      \\
		\hline
		thyroid			& 2715  &0.672 		&0.822   	&0.0410      \\
		&       &(0.833) 	&(0.918)   	&(0.0277)      \\
		\hline
		urogenital 		& 17112 &0.715 		&0.849   	&0.0363     \\
		system		&       &(0.825) 	&(0.914)   	&(0.0282)     \\
		\hline
	\end{tabular}
	\caption{Tissue-Specific Test. The first column is the 13 tissue names and we range them in alphabetical order. The second column is the ground true $\text{IC}_{50}$ values for each tissue. The last two columns are the $R^2$ and $R_p$ our model achieved by training using all the other tissue data. The number in the bracket is the result for the validation set.}
	\label{table:tissue_specific}
\end{table}

%% statistical significant difference
%% tissue blind
%% discuss the tissue distance

\subsection{Cell Lines Features Impacts}
In GDSC, the $735$ features for cell lines after preprocessing belongs to $310$ gene mutation states, and $425$ copy number variations. Since different laboratories may use different methods to extract the features for cell lines, in reality, it is not easy to have the complete $735$ features for all cells. More possibly, researchers may have smaller and different features group for cell lines. It will be attractive if our tCNNS can have good performance with fewer features for cell lines. In this part, we test tCNNS performance with a different number of features for cell lines. All the results in this part are shown in Figure \ref{fig:mutation_number}.

%%Assert that 300 mutation states would be sufficient to achieve comparable performance instead of using the full 735 states.

\begin{figure}[h]
	\centering
	\caption{Sensitivity to the number of features. The x-axis is the number of mutation states used for cells in the experiments, and the y-axis is the performance we achieved.}
	\label{fig:mutation_number}
\end{figure}

%%Discuss the advantage of using fewer mutations: fewer required mutations indicate fewer experiments to determine the necessary mutations only, therefore costing less budget and less time for experimentalists. Compare to previous works. menden2013machine, the characterization of copy number variation and micro satellite will no longer be required, thus saving a lot of efforts.

%% Discuss the result for less number of features 
%% 1 gene mutation states, 
%% 2 coding variants, 
%% 3 copy number variation.

\subsection{Biological Meaning v.s Statistical Meaning}
%% figure for ordering
tCNNS takes the one-hot representation of the SMILES format as the features for drugs. Initially, in the one-hot representation of the SMILES format, each row represents a symbol, and each column represents a position in the SMILES format, which is left aligned. For human researchers, the SMILES format is a well-defined concept with biological meaning. However, we think tCNNS may lack the ability to comprehend the biological meaning of the SMILES format and it instead relies on the statistical pattern inside the data. To verify this hypothesis, we modify the one-hot representation of the SMILES format in three ways. Firstly, we randomly shuffle the order of the symbols, which equals shuffling the rows in the one-hot representation. Secondly, we cut the SMILES format into two pieces and switch their positions, which equals shifting the columns in the one-hot representation. Thirdly, we shuffle the positions in the SMILES format, which equals shuffling the columns in the one-hot representation. In the last two ways of the modification, the biological meaning of SMILES is corrupted. We repeat the experiments in the three settings for $10$ times respectively and compare the results to the result using the SMILES format without any modification as the benchmark. The comparison is shown in supplementary Figure 5.

Since the computation of different channels (the rows in the one-hot representation of SMILES format) are independent of each other in the convolution neural network, we think the result should keep the same in the first modification with the benchmark. In the second modification, shifting the SMILES will only change the three column in the one-hot representation: the two ends columns and the column where we start the shifting. Meanwhile,  other columns keep the same. Thus, we also think that the result in the second modification should keep the same with the benchmark. The experimental results do support our predictions in the first two modifications. What surprised us is that the performance also keeps similar in the third modification. We used to think that shuffling the positions in the SMILES format will deteriorate the result because it corrupted the structure of the SMILES format. This stability means that tCNNS actually does not capture the biological meaning of the SMILES format for drugs, and it relies on the statistical patterns inside the SMILES format, cell line features, and the $\text{IC}_{50}$ values.

%% move to early position
\subsection{Eliminating Outlier} 

In the last column in supplementary Figure 5, we compared the results of tCNNS with the results of benchmark \cite{menden2013machine}. As GDSC has changed in recent years, we can not use the same data as \cite{menden2013machine}. In the experiment, we follow the method introduced in \cite{menden2013machine} and apply it to current data. We use PaDEL(version 2.1.1) to extract $778$ features for each drug. For cell lines, we use the same $735$ features, instead of the 157 features in the old version of GDSC used in \cite{menden2013machine}. 

\begin{figure}[h]
	\iffalse
	\begin{subfigure}{0.23\textwidth}
		\includegraphics[width=\linewidth]{imgs/drug_padel_dot.eps}
		\caption{drugs in PaDEL space}
	\end{subfigure}\hfill
	\begin{subfigure}{0.23\textwidth}
		\includegraphics[width=\linewidth]{imgs/drug_cnn_dot.eps}
		\caption{drugs in CNN space}
	\end{subfigure}\hfill
	\begin{subfigure}{0.23\textwidth}
		\includegraphics[width=\linewidth]{imgs/cell_mut_dot.eps}
		\caption{Cells in mutation space}
	\end{subfigure}\hfill
	\begin{subfigure}{0.23\textwidth}
		\includegraphics[width=\linewidth]{imgs/cell_cnn_dot.eps}
		\caption{Cells in CNN space}
	\end{subfigure}
	\fi
	\caption{Visualization of drugs and cells in the high-dim space. a) Drugs in PaDEL space(778 dims), b) Drugs in CNN space(420 dims), c) cells in mutation space(735 dims), d) cells in CNN space(1680 dims)}
	\label{fig:drug_cell_tsne}
\end{figure}

To compare the features extracted using PaDEL and features extracted from the SMILES format using CNN, we visualize the distribution of the drugs in different feature spaces to check the difference. When drawing the distribution of drugs using CNN, we use the output of the last layer of CNN tranche for drugs. The reason is that in deep neural network, the fully connected layer is responsible for regression analysis, and CNN is for extracting the high-level features from the drugs features. The input data for the fully connected network is the output of CNN tranche. So drawing the distribution of drugs using the output of the last layer in CNN is more reasonable. 

We also compared the distribution of cell lines in features space of GDSC, and in the output space of the last layer in CNN respectively. The visualization tool we used is t-SNE\cite{maaten2008visualizing}, which is widely used to visualize high dimensional data in deep learning. 

The visualization results are shown in Figure \ref{fig:drug_cell_tsne}. We can see that CNN can distribute the drugs and cell lines more uniformly than features using PaDEL and features of GDSC. For drugs, we see that there are seven outliers in the PaDEL features space. These seven outliers are the only seven drugs that are composed of multiple parts. We show the structure of those drugs in supplementary Figure 6.

\section{Conclusion}
\label{sec:conclusion}
In this paper, we came up with a model tCNNS for phenotypic screening between cancer cell lines and anti-cancer drugs. tCNN is tested on a new version of GDSC with more data compared to previous works. It achieved a much better coefficient of determinant and Pearson correlation than previous works and made predictions for missing values in GDSC with trustful evidence. tCNNS can also converge with a very small set of training data and fewer features for cancer cell lines, which is economically efficient. tCNNS took SMILES as the input data for drugs, and this eliminated the outlier problem in previous works in which drugs fingerprint are used as the features. tCNNS performs better in drugs repurposing than drug discovery, this may be a future research topic.

%Improving Prediction of Phenotypic Drug Response on Cancer Cell Lines Using Novel Features and Deep Learning. Emphasize more data (GDSC v6 VS v2). Emphasize novelty: SMILES features. Emphasize good result: deep learning.
\section{Competing interests}
The authors declares that they have no competing interests.
\section{Declaration}
The authors declares that they have provided the code and data public accessible.

\appendix

\end{document}